\documentclass[amsmath,amssymb,prd,nofootinbib]{revtex4}

\usepackage{graphicx}
\usepackage{dcolumn}
\usepackage{bm}
\usepackage{epsf}
\usepackage{hyperref}
\usepackage{amsfonts}

\begin{document}

\title{Self Force Orbit-Integrated gravitational waveforms for E(I)MRIs \\
(Talk given at the $9^{\rm th}$ LISA Symposium, 21--25 May 2012, BnF--Paris)}

\author{Lior M.~Burko$^{1,2}$ and Kristen A.~Lackeos$^2$} 

 \affiliation{$^1$ Universit\'e d'Orl\'eans, Observatoire des Sciences de l'Univers en Region Centre, LPC2E Campus CNRS, 45071 Orl\'eans, France\\
 $^2$ Department of Physics, University of Alabama in Huntsville, Huntsville, Alabama 35899, USA
 }
 
 \date{May 22, 2012}
 
\begin{abstract} 

We present the first orbit--integrated self force effects  for an IMRI or EMRI source, specifically the effects of its conservative piece on the orbit and on the waveform. We consider the quasi--circular motion of a particle in the spacetime of a Schwarzschild black hole, find the orbit and the corresponding gravitational waveform, and discuss the importance of the conservative piece of the self force in detection and parameter estimation. We also show the effect of the conservative piece of the self force on gauge invariant quantities, specifically $u^t$ as a function of the angular frequency $\Omega$. For long templates the inclusion of the conservative piece is crucial for gravitational--wave astronomy, yet may be ignored for short templates with little effect on detection rate. 
\end{abstract}

\maketitle


An important source of gravitational waves for space borne detectors are E(I)MRIs, extreme (intermediate) mass ratio inspirals. Such sources will allow us to test directly the Kerr hypothesis, and allow us to map the spacetime surrounding black holes. Moreover, the detection of E(I)MRIs will allow us to determine the mechanisms that shape stellar dynamics in galactic nuclei with unprecedented precision (\cite{Amaro:2012} and references cited therein). 

The orbits of E(I)MRIs are typically highly relativistic, and exhibit exciting phenomena, e.g.~extreme periastron and orbital plane precessions. Because the orbital evolution time scale (``radiation reaction time scale'') is much longer than the orbital period(s), over short time scales the orbit is approximately geodesic, yet on long time scales it deviates strongly from geodesic motion of the background. Instead, the smaller objects moves along a geodesic of a perturbed spacetime. Equivalently, one may construe the orbit as an accelerated, non-geodesic motion in the spacetime of the unperturbed central object, where the acceleration is caused by the self force  of the smaller object (\cite{Poisson-Pound-Vega}). We undertake here the latter approach. 

We consider here a point particle of mass $\mu$ that moves in a quasi circular orbit (i.e., an orbit that would be circular but for the radiation reaction effects) in the space-time of a Schwarzschild black hole of mass $M$, under the assumption that $\mu\ll M$. Specifically, we take here $\mu=10^{-2}\,M$, and the orbit starts at $r_0=10\,M$ down to close to the ISCO at $r=6\,M$. The orbital evolution is driven by the particle's self force. In practice we use the self force found for circular orbits in the Lorenz gauge and interpolate (using the least number of necessary terms) the tabulated values to obtain smooth functions for the self force to the accuracy given in \cite{barack-sago:2007}. Specifically, we expand the components of the self force as follows:
\begin{eqnarray*}
f^t_{r\le 8\,M}&=&-
\frac{1}{\sqrt{1-\frac{3M}{r}}\left(1-\frac{2M}{r}\right)} \; \bigg(\frac{M}{r}\bigg)^5\; \bigg(\frac{\mu}{M}\bigg)^2\;
\Bigg[a^-_0+a^-_1\,\frac{M}{r}+a^-_2\,\bigg(\frac{M}{r}\bigg)^2+a^-_3\,\bigg(\frac{M}{r}\bigg)^3 + \cdots \Bigg]
\nonumber\\
f^t_{r\ge 8\,M}&=&-
\frac{32}{5}\,\frac{1}{\sqrt{1-\frac{3M}{r}}\left(1-\frac{2M}{r}\right)} \;\left(\frac{M}{r}\right)^5\;\left(\frac{\mu}{M}\right)^2\;
\left[ {\rm PN}_{5.5}+\left(a^+_6+a^+_{6L}\,\ln\frac{M}{r}\right)\left(\frac{M}{r}\right)^6 + \cdots \right]
\nonumber\\
f^r_{r\le 8\,M}&=&
\bigg(1-\frac{2M}{r}\bigg)\; \bigg(\frac{M}{r}\bigg)^2
\bigg(\frac{\mu}{M}\bigg)^2\;
\Bigg[b^-_0+b^-_1\,\bigg(1-\frac{6M}{r}\bigg)+b^-_2\,\bigg(1-\frac{6M}{r}\bigg)^2+b^-_3\,\bigg(1-\frac{6M}{r}\bigg)^3 + \cdots\Bigg]\nonumber\\
f^r_{r\ge 8\,M}&=& \bigg(\frac{M}{r}\bigg)^2
\bigg(\frac{\mu}{M}\bigg)^2\;
\Bigg[b^+_0+b^+_1\,\frac{M}{r}+b^+_2\,\bigg(\frac{M}{r}\bigg)^2
+b^+_3\,\bigg(\frac{M}{r}\bigg)^3 + \cdots\Bigg]\nonumber
\end{eqnarray*}
where ${\rm PN}_{5.5}$ stands for the standard $\frac{11}{2}$--post--Newtonian expression (converting Eq.~(3.1) in \cite{tanaka:1996} from luminosity to $f^t$). Fitting the free parameters, we find the values appearing in Table 1.
\begin{table}[htdp]
\caption{The fit parameters for the self force. These parameters reproduce the accuracy of \cite{barack-sago:2007} to all significant figures for all data points. Our results for $a^+_{6,6L}$ are very crude predictions for the corresponding PN parameters, as our fit ignores all higher--order terms.}
\begin{center}
\begin{tabular}{|c|c||c|c||c|c||c|c||}\hline\hline
$a^-_0$ & 4.57583 & $a^+_6$ & 331.525 & $b^-_0$ & 1.32120 & $b^+_0$ & 1.999991\\
\hline
$a^-_1$ & 31.8117 & $a^+_{6L}$ & -2081.57 & $b^-_1$ & 1.2391 & $b^+_1$ & -6.9969\\
\hline
$a^-_2$ & -267.250 & & & $b^-_2$ & -1.297 & $b^+_2$ & 6.29\\
\hline
$a^-_3$ & 1049.27 & & & $b^-_3$ & 1.07 & $b^+_3$ & -24.6\\
\hline\hline
\end{tabular}
\end{center}
\label{default}
\end{table}%
Finally, the last remaining non--zero component of the self force ($u^{\phi}$) is found from the condition that the self force must satisfy $f^{\rm SF}_{\mu}\,u^{\mu}=0$, where the 4--velocity is taken for the circular geodesic orbit for which the self force was calculated. 

\begin{figure}[!h]
\includegraphics[width=10.0cm,angle=0]{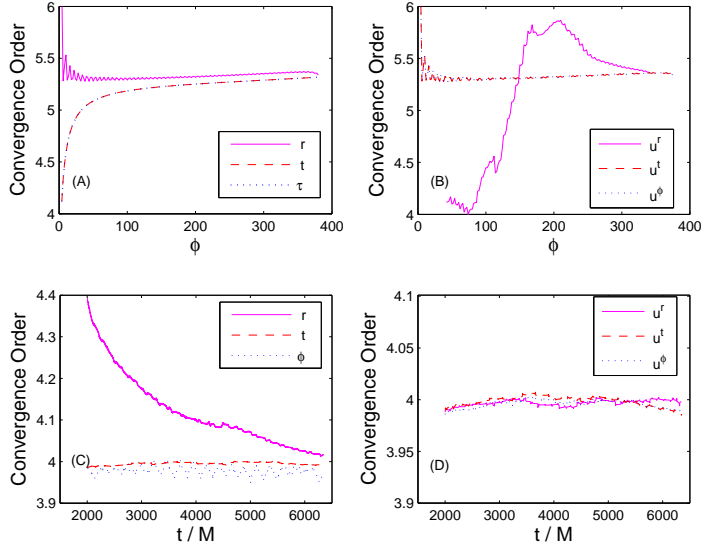}
\caption{Convergence tests for the two codes. Top: the 4-position (A) and 4-velocity (B) for the osculating code as functions of the azimuthal angle $\phi$. Bottom: the 4-position (C) and 4-velocity (D) for the direct code as a function of the time $t$.}
\label{fig1}
\end{figure}

\begin{figure}[!h]
\includegraphics[width=10.0cm,angle=0]{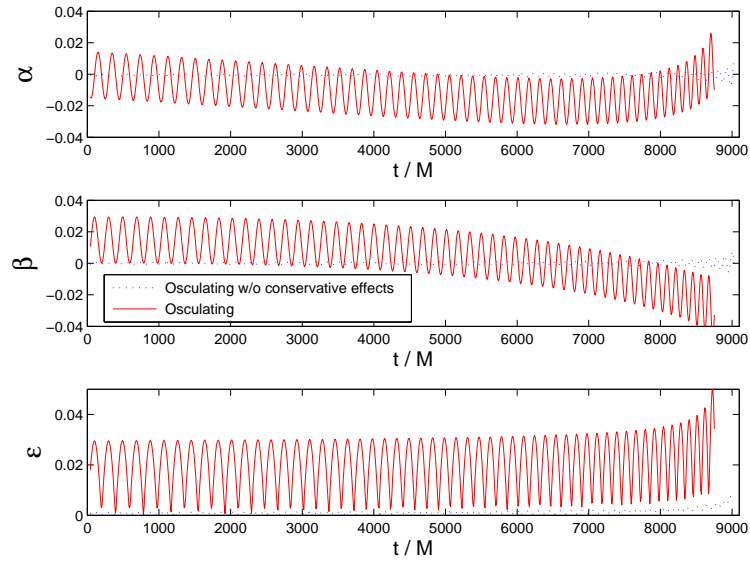}
\caption{The osculating--code variables $\alpha$ and $\beta$ as functions of the time $t$ (upper two panels), and the effective eccentricity $\epsilon$ of the orbit (lower panel) as a function of $t$ in the osculating case.}
\label{fig2}
\end{figure}

We evolve the orbit using three different codes. First, we use the energy balance approach to evolve the orbit. This approach ignores conservative effects on the orbital evolution and correspondingly on the waveforms, and can be done without the self force: fluxes of otherwise conserved qualities to infinity and down the event horizon are used to update the particle's constants of motion. As expected, we show in \cite{burko:2012} that the orbit and the resulting waveforms are identical in the energy balance approach and in our self force calculations, when the conservative piece of the self force is turned off (i.e., when we take $b_i^A=0$.) We used two independent codes for the computation of the orbital evolution with the self force. First, we used the method of osculating geodesics (specifically eqs. (43)--(47) in \cite{pound:2008}). We also integrate the equations of motion directly, i.e., solve directly for $u^{\mu}$ from the equation of motion $u^{\beta}\,\nabla_{\beta}u^{\alpha}=\mu^{-1}\,f^{\alpha}_{\rm SF}$ (``Newton's second law,'' with covariant differentiation compatible with the background metric) and integrate its solution to find $x^{\mu}$. Both codes are numerically stable and convergent. Specifically, the osculating code converges with 5$^{\rm th}$ order, and the direct code converges with $4^{\rm th}$ order (Fig.~\ref{fig1}). 

The integration using the method of osculating geodesics cannot keep the value of the eccentricity as precisely zero. As both variables $\alpha$ and $\beta$ (see \cite{pound:2008} for definitions) are dynamical, the eccentricity $\epsilon$ must evolve along the orbit too. This behavior is shown in Fig.~\ref{fig2}. Interestingly, the inclusion of the conservative piece of the self force amplifies the resulting effective eccentricity.

\begin{figure}[!h]
\includegraphics[width=10.0cm,angle=0]{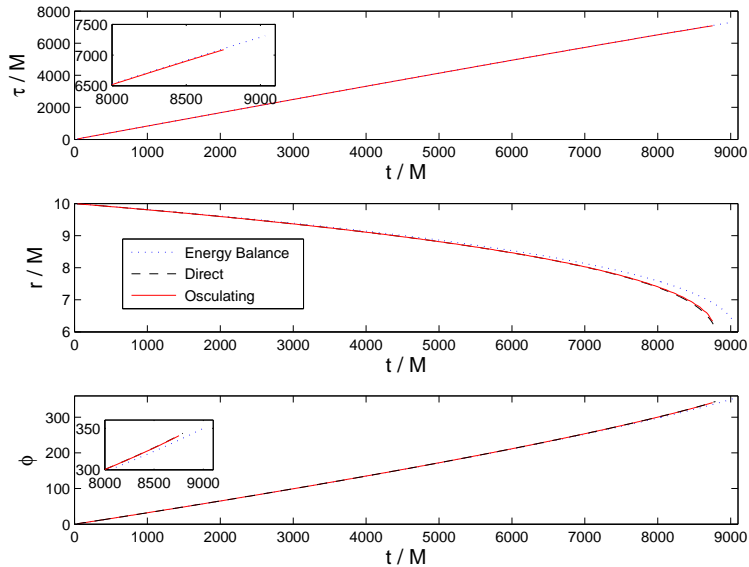}
\caption{The Orbit. The 4--position for the three orbital evolution codes: energy balance (dotted), direct evolution (dashed), and osculating code (solid).}
\label{fig3}
\end{figure}

\begin{figure}[!h]
\includegraphics[width=10.0cm,angle=0]{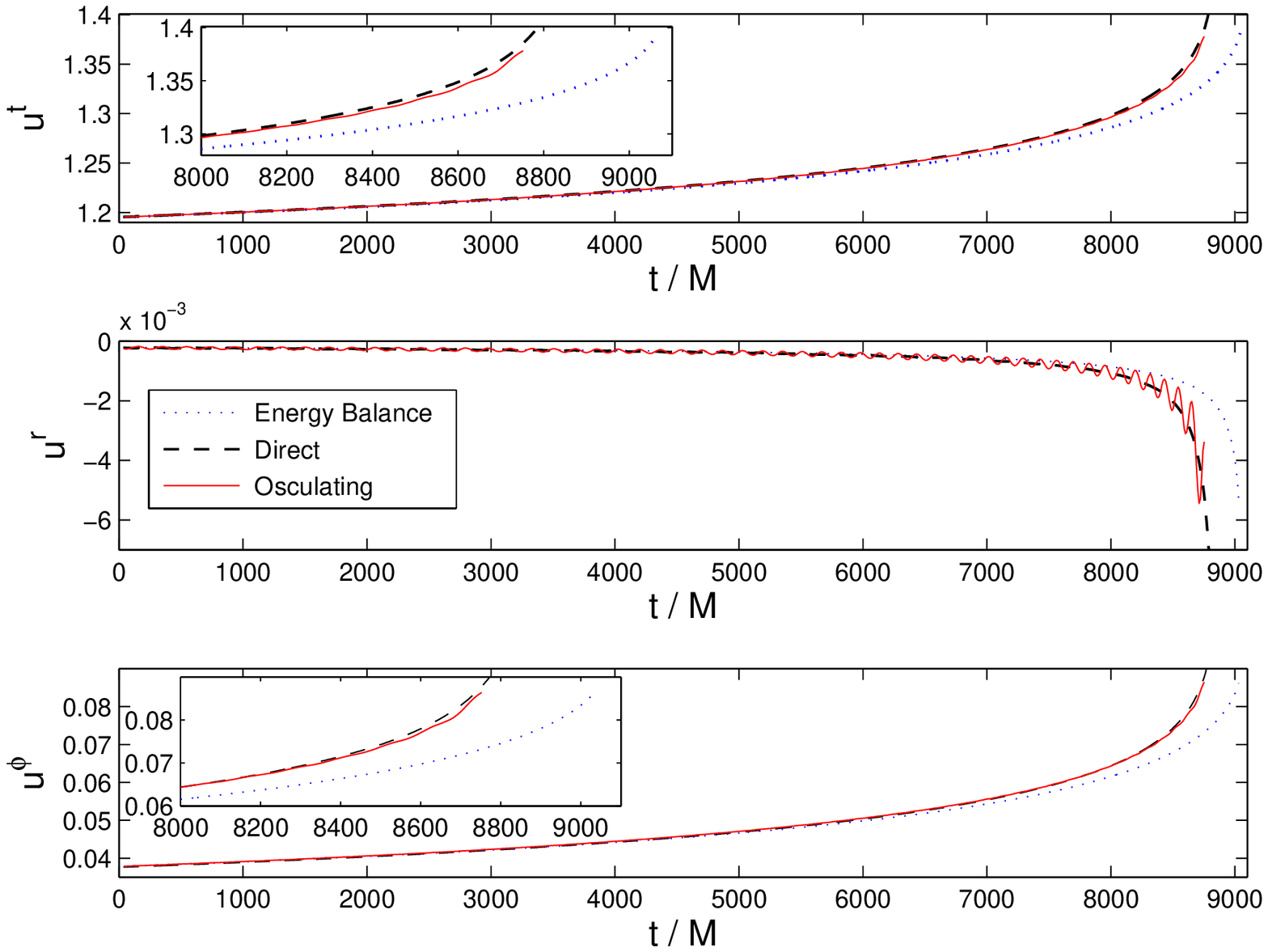}
\caption{The Orbit. The 4--velocity for the three orbital evolution codes: energy balance (dotted), direct evolution (dashed), and osculating code (solid).}
\label{fig3b}
\end{figure}

\begin{figure}[!h]
\includegraphics[width=10.0cm,angle=0]{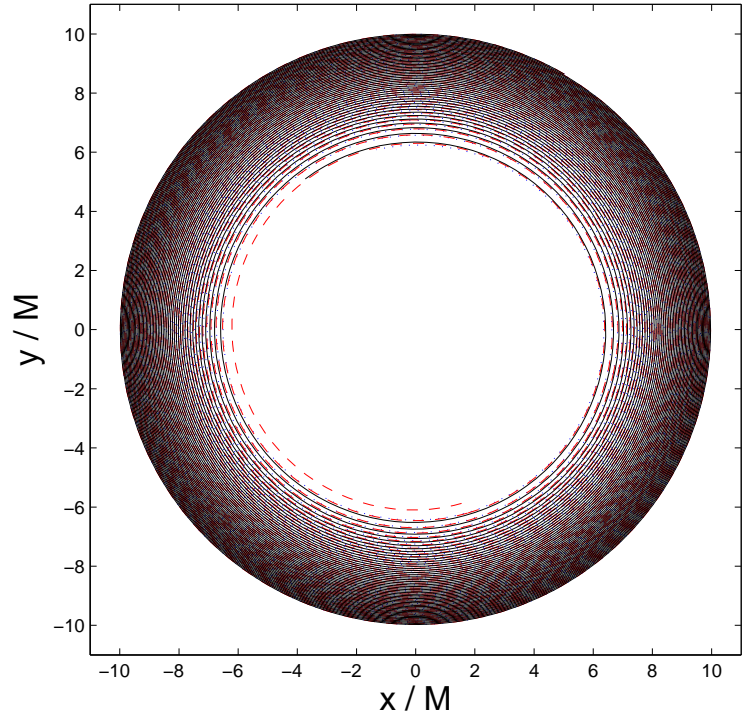}
\caption{The Orbit. The shape of the orbit for the three orbital evolution codes: energy balance (dotted), direct evolution (dashed), and osculating code (solid).}
\label{fig3c}
\end{figure}

\begin{figure}[!h]
\includegraphics[width=10.0cm,angle=0]{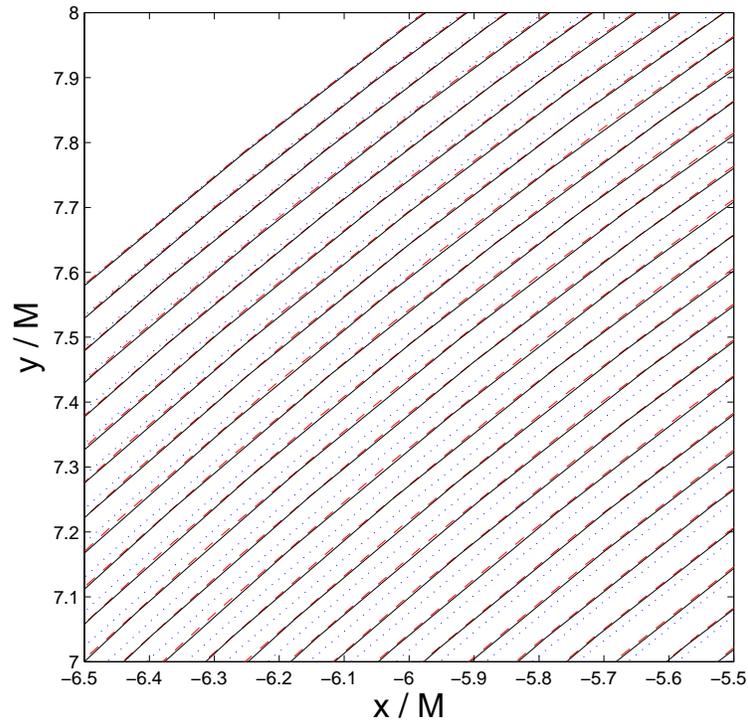}
\caption{The Orbit. A small portion of the orbit is magnified to show detail.}
\label{fig3d}
\end{figure}

The orbit is displayed in Figs.~\ref{fig3} and \ref{fig3b} and Figs.~\ref{fig3c} \ref{fig3d} for the three codes. Notably, the two independent self force codes reproduce the orbit to high level of agreement, with a difference much smaller than the difference between either and the orbit generated in the energy balance approach. This difference is attributed to the effect of the conservative piece of the self force.

\begin{figure}[!h]
\includegraphics[width=10.0cm,angle=0]{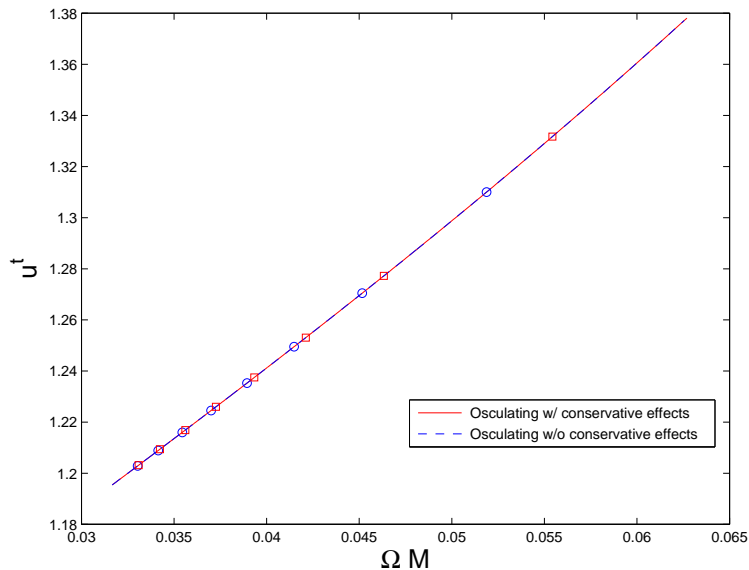}
\caption{The gauge invariant figure of $u^t$ as a function of the angular frequency $\Omega$. The curve without (dashed, $\circ$) and with the conservative effects  (solid, $\square$) are shown, together with equal-$t$ spacing marks starting at $t=1,500\,M$ in increments of $1,000\,M$. The two curves are indistinguishable to the numerical accuracy of our computation 
(using the osculating code in both cases).}
\label{fig4}
\end{figure}

The orbit, of course, is a gauge dependent quantity. Indeed, the position vector changes trivially under gauge transformations, $x^{\mu}\to x^{\mu}+\xi^{\mu}$. We can, however, create gauge invariant quantities with a specific gauge choice (in our case, the Lorenz gauge), and then those quantities -- by virtue of their gauge invariance -- are guaranteed to remain unchanged in any other gauge. Two independent gauge invariant quantities are $u^t$ (``gravitational redshift") and the angular frequency $\Omega$ (\cite{detweiler:2008}). In Fig.~\ref{fig4} we plot $u^t$ as a function of $\Omega$ with and without the conservative piece of the self force. Notably, to the accuracy of our numerical computation the two curves overlap. That is, we find that $u^t$ as a function of $\Omega$ is insensitive to the conservative piece of the self force. This conclusion implies that when an actual data stream is used and this gauge invariant figure is plotted, one may use a simplified radiation--reaction scheme, that does not include the conservative effects in its analysis. 

There is, however, an aspect of the gauge invariant figure that is sensitive to the conservative effects, specifically the speed with which the data point moves along the curve. The way the conservative effects are manifested in the gauge invariant plot is not is the shape of the curve, but in the time it takes the signal to move along it. 

\begin{figure}[!h]
\includegraphics[width=17.0cm,angle=0]{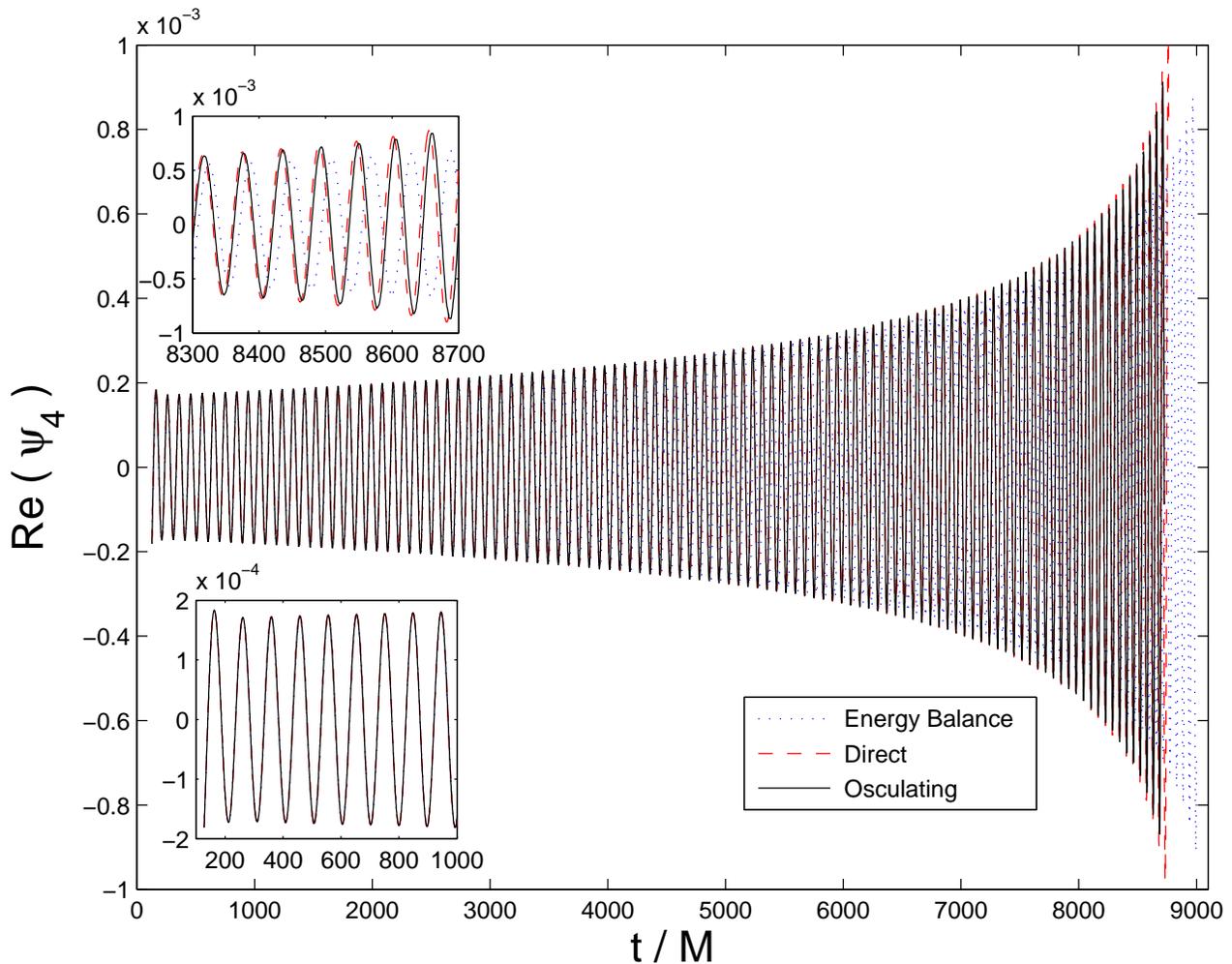}
\caption{The real part of the waveform $\psi_4$ when the conservative piece of the self force is included (solid for the osculating orbit method, and dashed for the direct integration method), and using the energy balance approach (dotted). The three waveforms are in phase at the beginning of the waveforms, but the former two dephase with time from the latter (and to a much smaller extent from each other).}
\label{fig5}
\end{figure}

After we obtain the orbit we used it in order to evolve the linearized Einstein equations, using a code for the sourced Teukolsky equation with hyperboloidal slicing (\cite{zenginoglu:2011}). We present the waveforms in Fig.~\ref{fig5}. We show the three waveforms, for the energy balance approximation --- which is equivalent to the self force waveform with the conservative piece turned off --- and the two independent waveforms obtained from the direct approach and from the osculating geodesics approach that use the full self force, including its conservative piece. All three waveforms start in phase at early times. The two full self-force waveforms are dephased from the energy balance waveforms by much more than from each other. Specifically, the total commutative dephasing of either waveform from the energy balance waveform over the entire orbit from $r_0=10\,M$ down to near the ISCO, is $\,\Delta\phi=8.4\pm 0.4\;{\rm rad}$. This dephasing corresponds to about $4/3$ of a cycle, compared with the $54.3$ cycles the particle makes in the energy balance case. 

We next study the importance of the conservative piece of the self force for the waveforms. Specifically, we take a window of duration $L$ from the end of the chirp part of the energy balance waveform. and find its overlap integral with the full self force waveform, which in this case models the actual data stream. As expected, the longer $L$, the smaller the overlap integral. When we take $L=L_{\rm threshold}=816.6\,M$ we find that the overlap integral is reduced to $0.96$, which corresponds to a loss of $10\%$ in the event rate.  Therefore, for $L\lesssim L_{\rm threshold}$ ignoring the conservative piece of the self force does not lead to a significant loss of accuracy in detection of E(I)MRI events. However, longer stretches of data, $L\gtrsim L_{\rm threshold}$ leads to a significant decrease in event rates, so that the inclusion of the conservative piece of the self force is important. Further detail of this work appears in \cite{burko:2012}.



The authors are indebted to Gaurav Khanna for discussions and for use of his numerical code. 
This work was supported by a NASA EPSCoR RID grant and by NSF grants PHY--0757344 and DUE--0941327. LMB is grateful to Alessandro Spallicci for hospitality.


\bibliography{/Users/Lior/Documents/Papers/Lisa_9/MyBibli}
\bibliographystyle{/Users/Lior/Documents/Papers/Lisa_9/asp2010}

 \end{document}